\documentclass[aps,pra,floatfix,superscriptaddress,showpacs,twocolumn]{revtex4}
\usepackage{amsmath}
\usepackage{graphicx}
\def\prn#1{{\left(#1\right)}}

\def\prn#1{{\left(#1\right)}}

\begin{document}
\title{Measurement of excited-state transitions in cold calcium atoms by direct femtosecond frequency-comb spectroscopy}
\author{J.E. Stalnaker}
\affiliation{National Institute of Standards and Technology, Time
and Frequency Division, MS 847 Boulder, CO 80305}
\author{Y. Le Coq}
\affiliation{National Institute of Standards and Technology, Time
and Frequency Division, MS 847 Boulder, CO 80305}
\author{T.M. Fortier}
\affiliation{Los Alamos National Laboratory, P-23 Physics Division
MS H803, Los Alamos, NM 87545}
\author{S.A. Diddams}
\affiliation{National Institute of Standards and Technology, Time
and Frequency Division, MS 847 Boulder, CO 80305}
\author{C.W. Oates}
\affiliation{National Institute of Standards and Technology, Time
and Frequency Division, MS 847 Boulder, CO 80305}
\author{L. Hollberg}
\affiliation{National Institute of Standards and Technology, Time
and Frequency Division, MS 847 Boulder, CO 80305}

\date{\today}
\begin{abstract}
We apply direct frequency-comb spectroscopy, in combination with
precision cw spectroscopy, to measure the ${\rm 4s4p} \;^3P_1
\rightarrow {\rm 4s5s} \; ^3S_1$ transition frequency in cold
calcium atoms.  A 657 nm ultrastable cw laser was used to excite
atoms on the narrow ($\gamma \sim 400$ Hz) ${\rm 4s^2} \;^1S_0
\rightarrow {\rm 4s4p} \;^3P_1$ clock transition, and the direct
output of the frequency comb was used to excite those atoms from the
${\rm 4s4p} \; ^3P_1$ state to the ${\rm 4s5s} \; ^3S_1$ state.  The
resonance of this second stage was detected by observing a decrease
in population of the ground state as a result of atoms being
optically pumped to the metastable ${\rm 4s4p} \;^3P_{0,2}$ states.
The ${\rm 4s4p} \; ^3P_1 \rightarrow {\rm 4s5s} \; ^3S_1$ transition
frequency is measured to be $\nu = 489\,544\,285\,713(56)$~kHz;
which is an improvement by almost four orders of magnitude over the
previously measured value. In addition, we demonstrate spectroscopy
on magnetically trapped atoms in the ${\rm 4s4p} \;^3P_2$ state.
\end{abstract}


\maketitle

By providing a clockwork to divide optical frequencies down to
countable rf frequencies, stabilized femtosecond frequency combs
have made possible absolute frequency metrology of cw lasers used to
probe atomic transitions at unprecedented levels of accuracy
\cite{oskay06,ludlow06,margolis04,peik04,fischer04}. In addition,
direct frequency-comb spectroscopy has been performed in a variety
of systems including two-photon and single photon transitions
\cite{marian05,gerginov05,witte05,fortier06c}.  Here we extend the
techniques developed using direct frequency comb spectroscopy to the
study of transitions between excited states by performing step-wise
spectroscopy in combination with a cw laser. Significantly, we
exploit the frequency comb's wavelength versatility to directly
excite transitions between excited states of cold calcium (Ca) atoms
at 612~nm and 616~nm, eliminating the need for lasers operating at
these wavelengths.  By utilizing these techniques in conjunction
with cold atoms we are able to surmount the low power limitations
and have achieved an uncertainty of 56 kHz, despite having only
$\approx 75$ nW of power in the resonant mode.

The frequency comb is generated by an octave-spanning femtosecond
mode-locked laser based on Ti:sapphire with a repetition rate of
$\approx 1$ GHz and is described in detail in Ref.\
\cite{fortier06a}. The frequency comb provides $\approx 10^5$
optical modes that are related to two rf frequencies by
\begin{align}
\nu_n=f_0+n \: f_{{\rm Rep}},
\end{align}
where n is the mode number, $f_0$ is the carrier-envelope offset
frequency and $f_{\rm Rep}$ is the repetition rate of the laser. The
carrier-envelope offset frequency is stabilized by use of the
self-referencing method with a $f$-to-$2f$ interferometer
\cite{jones00}.  The repetition rate is phase locked to a
synthesizer referenced to a hydrogen maser. The stabilization of the
comb in this manner results in the stabilization of all of the
optical frequencies of the comb with fractional uncertainties below
the $10^{-12}$ level in one second.

\begin{figure}[h]
\centerline{\includegraphics[width=2.25 in]{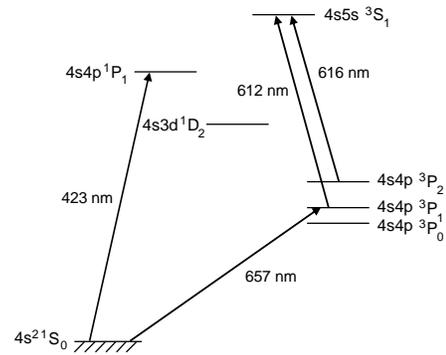}}
\caption{Low-lying energy levels of Ca.} \label{fig CaEnergy}
\end{figure}

The comb provides $\approx200$ nW of power per optical mode at 612
nm. Because light in the wings of the optical spectrum results from
continuum generation within the Ti:S laser crystal the resulting
spatial mode is not Gaussian  \cite{fortier06a}. Consequent
difficulties in coupling this light into single mode fiber led us to
to produce 612 nm light via broadening in highly nonlinear fiber
that produced a comparable level of light at 612 nm but with a clean
Gaussian mode. The light was then filtered using a 612 nm bandpass
interference filter with a passband of 3 nm and sent through $\sim
20$ m of optical fiber to the Ca atoms.  Due to losses in the fiber
coupling and in the optics used to deliver the light to the atoms,
approximately $70 \, {\rm \mu W}$ of light over a 3 nm bandwidth was
delivered to the atoms. Using an optical spectrum analyzer we
estimate there was approximately 75 nW per optical mode at the
resonant wavelength.

An overview of the experiment is shown in Fig.\ \ref{fig
Ca612Setup}. The experiment utilized an apparatus built for an
optical frequency standard based on the ${\rm 4s^2} \; ^1S_0
\rightarrow {\rm 4s4p} \; ^3P_1 \: M=0$ clock transition in
$^{40}{\rm Ca}$ that is described in detail in Ref.\cite{oates99}.
The atoms were laser cooled and trapped on the ${\rm 4s}^2 \: ^1S_0
\rightarrow {\rm 4s4p} \: ^1P_1$ transition at 423 nm (Fig.\
\ref{fig CaEnergy}) for 2.5 ms. The resulting sample consisted of
$\approx 6 \times 10^7$ atoms at a temperature of 2 mK. The cold
ground-state atoms were excited on the ${\rm 4s^2} \; ^1S_0
\rightarrow {\rm 4s4p} \; ^3P_1 \: M=0$ transition, which has a
natural linewidth of 374 Hz \cite{degenhardt05}, using linearly
polarized resonant light at 657 nm consisting of a $\pi$ pulse with
a duration of $2.5 \: {\rm \mu s}$. This excitation selects atoms
near zero velocity. Approximately 15 $\%$ of the atoms were excited
to the $^3P_1$ state. After a delay of $40 \: {\rm \mu s}$ the
number of atoms in the ground state was probed with a $50\: {\rm \mu
s}$ pulse of 423 nm light resonant with the ${\rm 4s^2} \; ^1S_0
\rightarrow {\rm 4s4p} \; ^1P_1$ transition, and the subsequent
fluorescence was detected with a photomultiplier tube. The entire
cooling-trapping-excitation-probe cycle was repeated every 2.6 ms.

\begin{figure}
\centerline{\includegraphics[width=2.75 in]{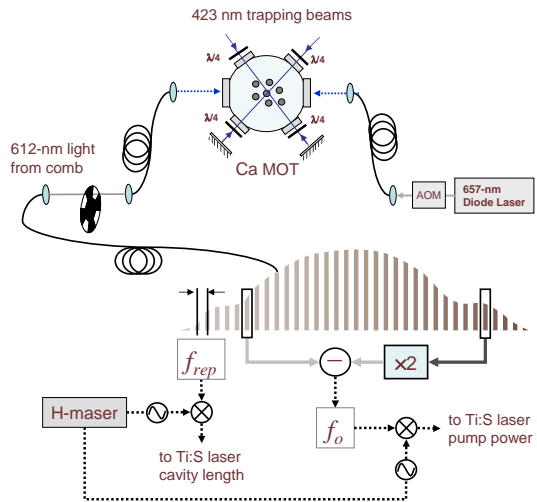}}
\caption{Schematic of experimental setup for measurement of $^3P_1
\rightarrow ^3S_1$ transition.} \label{fig Ca612Setup}
\end{figure}

The repetition rate of the frequency comb was scanned over a range
of 250 Hz, corresponding to a change of 125 MHz in the optical
frequency of the mode used to excite the $^3P_1 \rightarrow ^3S_1$
transition. When the comb light is resonant with the transition,
atoms are excited to the $^3S_1$ state, from which they decay to the
$^3P_J$ states with an approximate branching ratio of $5:3:1$ for
the $J=2,1,0$ states, respectively. Consequently, roughly $2/3$ of
the atoms decay to the long lived metastable $^3P_0$ and $^3P_2$
states. The decay to these states provides a loss mechanism for the
trapped atom number.  While atoms in the $^3P_1$ state can decay
back to the ground state and be recaptured in the next loading
sequence, the atoms that decay to the meatstable $J=0,2$ states are
do not decay to the ground state. In steady state, this leads to a
decrease in the number of atoms detected with the probe pulse. The
612 nm light from the comb was modulated with a optical chopper
wheel operating with a 50 $\%$ duty cycle at a frequency of 10 Hz.
The modulation frequency was chosen to be slow enough so that the
612 nm light leak mechanism would yield a sufficient depletion of
the number of atoms, while still providing sufficient suppression of
low-frequency noise. The population transfer was then measured by
looking for changes in the probe fluorescence via lock-in detection.

\begin{figure}
\centerline{\includegraphics[width=2.75 in]{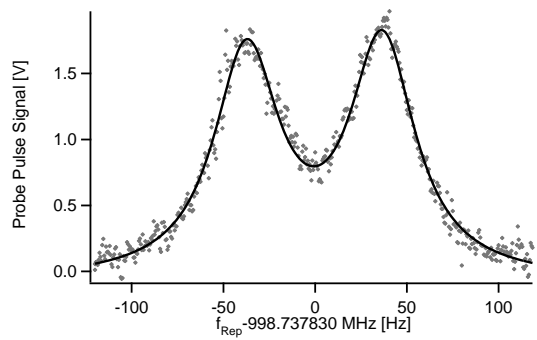}}
\caption{Typical fluorescence signal as the repetition rate of the
laser is scanned over the resonance.  The two peaks correspond to
transitions to the $M=\pm 1$ states of the $^3S_1$ state.  The
y-axis is the output of the lock-in amplifier; a change of +1 V
corresponds to a decrease in the population of the ground state of
$\approx 0.1 \: \%$. The scan time was $\approx 20$ minutes.}
\label{fig scan612}
\end{figure}

A typical signal is shown in Fig. \ref{fig scan612}.  The two peaks
correspond to the $M_J=\pm1$ components of the ${\rm 4s5s} \; ^3S_1$
state, which are split due to the residual field of the
magneto-optical trap. The $M_J=0$ component is not excited because
of the selection rule forbidding $M_J=0\rightarrow M_{J^\prime}=0$
transitions for states where $J^\prime=J$.  The 612 nm light is
roughly linearly polarized perpendicular to the 657 nm light.  We
attribute the difference in amplitudes of the two peaks to imperfect
polarization of the 612 nm light.

The data were fit to two Voigt profiles with the same widths.  This
lineshape was empirically seen to correctly describe the data.  The
actual lineshape is complicated by broadening due to the magnetic
field gradient in the center of the trap.  The average between the
center positions of the two peaks was used to extract the value of
the field-free resonance frequency.

In order to determine the optical frequency, the optical mode number
of the frequency comb resonant with the transition must be known.
This can be accomplished by varying the mode number, $n$, by a known
amount $\Delta n$ such that
\begin{align}
\frac{\delta f_{{\rm Rep}}}{f_{{\rm Rep}}} \ll \frac{\Delta n}{n^2},
\end{align}
where $\frac{\delta f_{{\rm Rep}}}{f_{{\rm Rep}}}$ is the fractional
uncertainty in the repetition rate of the resonance as determined
from the fit. We identified the optical mode number by adjusting the
repetition rate of the comb so as to shift the resonant mode number
by exactly 200 modes, corresponding to a change in the repetition
rate of the laser of $\approx 2$ MHz. Data were taken at four
different optical modes spanning 200 modes. Based on the consistency
of the data we were able to unambiguously identify the mode numbers.

\begin{figure}
\centerline{\includegraphics[width=2.75 in]{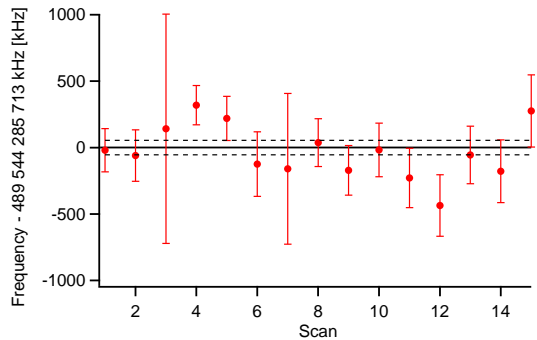}}
\caption{Measured resonance frequencies for the $^3P_1\rightarrow
^3S_1$ transition for 15 different scans. The dashed lines represent
the error on the mean.} \label{fig results}
\end{figure}

The extracted central frequencies of 15 data sets are shown in Fig.\
\ref{fig results}.  Each data set is the average of two scans of the
repetition rate over the resonance; one with increasing frequency
and one with decreasing frequency.  A difference between the scans
taken with increasing frequency and those taken with decreasing
frequency was observed.  The average difference was 40(26) kHz.  We
attribute this difference to slow changes in the alignment of the
microstructure fiber, leading to changes in the optical power as the
laser was scanned over the resonance.  As the power changes during
the scan, there is an effective shift in the center position of the
resonances. The amplitudes of the four peaks (two Zeeman states for
both the increasing and decreasing frequency scans) can be used to
determine the linear and quadratic drift in the power. This was done
for a data set exhibiting a large change in the power during the
course of the scan.  The coefficients thus determined were used to
ascertain the effect of the power changes on the measurement of the
peak centers. A shift of ~100 kHz was observed as a result of the
power drift. However, the effect of the drift canceled in the
average of the increasing frequency and decreasing frequency scans
at a level of 15 kHz.  We therefore assign an uncertainty of 15 kHz
due to power variations in the 612 nm light in the extracted
frequency.

There is also a systematic uncertainty resulting from the frequency
of the 657 nm cw light.  The 657 nm light was stabilized to a highly
stable optical cavity that has a drift rate of $\approx 5$ Hz/s,
corresponding to a drift of $\approx 2$ kHz over the course of the
scan. This drift results in a drift of the central velocity of the
atoms excited by the 657 nm light and therefore a drift in the
Doppler effect of the 612 nm transition. The result of the linear
drift on the extraction of the transition frequency is much less
than that due to the power fluctuations described above and is
further suppressed by the combination of the spectra taken with
increasing and decreasing repetition rates. Prior to a scan of the
612 nm light, the central frequency of 657 nm light was centered on
the zero-velocity class by use of Doppler-free spectroscopy to
within $\approx 50$ kHz. Because the center frequency of the 657 nm
light was recentered before each scan, we expect random variations
in this error.  We increased the statistical errors determined from
the fits by 50 kHz to account for this additional effect. After
including this effect, the spread of the data remains larger than
the uncertainties on the individual scans. We attribute this to a
possible motion of the trap during the scan, leading to shifts in
the average magnetic field to which the atoms are exposed. We
inflate the errors by a factor of 2.7 to get an appropriate chi
squared and take this scatter into account.  This gives an error on
the mean of 54 kHz.

We also considered the effect of ac Stark shifts due to the trapping
light.  Since the transition of interest is between two excited
states, the trapping light is far off resonance with both states.
Considering the possible states that could couple to the ${\rm
4s4p}\: ^3P_1$ state via a 423 nm photon, we estimate the ac Stark
shift of this state to be less than 20 Hz.  A 423 nm photon brings
the ${\rm 4s5s} \: ^3S_1$ state above the ionization limit, and the
ac Stark shift for the state is consequently negligible.  The small
optical power per mode and the 1 GHz separation of the optical modes
make ac Stark shifts from the non-resonant comb modes negligible.
Other systematic uncertainties that were considered are presented in
Table \ref{tab errorbudget}.

\begin{table}
\caption{Errors and the resulting uncertainty in the value of the
$^3P_1\rightarrow^3S_1$ transition. The type A uncertainty include
effects that average randomly, including the error due to starting
frequency of the 657 nm light, the statistical errors as determined
from the fits, and the additional scatter, which we attribute to
motion of the trap.} \label{tab errorbudget}
\begin{center}
\begin{tabular}{lc} \hline
\hline {\bf Effect} & {\bf Uncertainty [kHz]}
\\ \hline
Type A Uncertainty  & 54\\
Power changes in the 612-nm light & 15 \\
ac-Stark shifts from trapping light & $<0.02$  \\
Second-order Zeeman & $<0.010$ \\
Frequency Calibration of Comb & $<0.15$
\\\hline Combined Uncertainty & 56
\\\hline \hline
\end{tabular}
\end{center}
\end{table}

We arrive at a final value for the transition frequency of
\begin{align}
\nu\prn{^3P_1 \rightarrow ^3S_1}=489\: 544\: 285\: 713 (56)\: {\rm
kHz}.
\end{align}
This value is an improvement by almost four orders of magnitude over
the previously measured value as tabulated in Ref.\ \cite{sugar85}:
$\nu\prn{^3P_1 \rightarrow ^3S_1}=489\: 544\: 032(450)$~MHz.

\begin{figure}[h]
\centerline{\includegraphics[width=2.75 in]{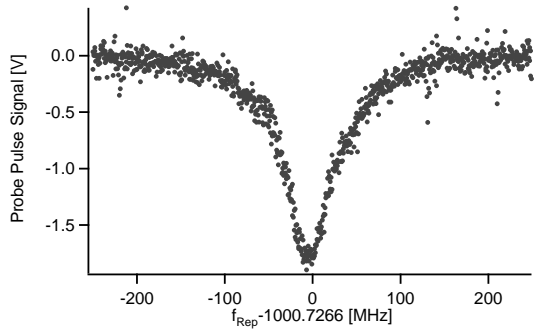}}
\caption{Fluorescence signal from the 423 nm probe as the repetition
rate of the frequency comb is scanned over the ${\rm 4s4p}\; ^3P_2
\rightarrow {\rm 4s5s}\; ^3S_1$ transition. A decrease in the
voltage corresponds to an increase in the ground-state population.}
\label{fig scan616}
\end{figure}

In addition to the measurement of the $^3P_1\rightarrow^3\!S_1$
transition, we were able to observe the ${\rm 4s4p}\;
^3P_2\rightarrow{\rm 4s5s}\; ^3S_1$ transition at 616 nm.   This
experiment was performed without directly exciting atoms to the
${\rm 4s4p }\: ^3P_2$ state.  Instead, the atoms were populated to
the $^3P_2$ state during the cooling and trapping process.  Atoms
excited to the $^1P_1$ state have a $\approx 10^{-5}$ probability to
decay to the $4s3d \; ^1D_2$ state. These atoms decay predominantly
to the $^3P$ states. Atoms that decay to the $^3P_2$ state can be
magnetically trapped in the magnetic field gradient due to the MOT
trapping coils.  By exciting these atoms on the ${\rm 4s4p} \; ^3P_2
\rightarrow {\rm 4s5s} \; ^3S_1$ transition at 616 nm we were able
to create a repumping mechanism that increases the number of atoms
in the ground state, as some of the atoms decay from the $^3S_1$
state to the $^3P_1$ state and then back to the ground state.  The
experiment is identical to that used for the measurement of the
$^3P_1 \rightarrow ^3S_1$ transition frequency, except that there is
no 657 nm $\pi$ pulse needed. The probe signal as the repetition
rate of the comb is scanned across the resonance is shown in Fig.\
\ref{fig scan616}.

Only states with $M=+1$ or $M=+2$ are magnetically trapped.  There
are three possible transitions from these states to the $M=0$ and
$M=+1$ Zeeman sublevels of the $^3S_1$ state.  Because of the
broadening due to the spatial distribution of the atoms in the
magnetic-field gradient, we were unable to resolve these
transitions. Consequently, we are unable to determine the Zeeman
shift of the transition and cannot calibrate the spatial
distribution of the atoms within the magnetic field gradient.  While
the atoms are distributed over a large volume within the magnetic
trap, the 616 nm laser beam intersects only a small region, about 1
mm diameter, of the trap, which is overlapped with the MOT.  In
order to estimate the magnetic field over the region of the probe
beam, we use the measurement of the Zeeman splitting observed on the
612 nm transition described above, which gives a residual field of
$6.3(3) \times 10^{-4}$ T. We estimate the magnetic field gradient
to be $60(12) \times 10^{-4}$ T/cm in the strong direction. Using
these values, and assuming a Gaussian spatial and velocity
distribution of atoms over the magnetic field region, gives an
estimate for the shift of the different transitions. We find the
transition frequencies shift by $-13$ MHz, $-9$ MHz, and 4 MHz for
the $M=1 \rightarrow M=0$, $M=2 \rightarrow M=1$, and $M=1
\rightarrow M=1$ transitions, respectively. If equal weights for the
three transitions are assumed and the different broadening of the
transitions is included, the effective shift of the resonance is
$-4$ MHz.  Due to uncertainties in the relative excitation rates for
the different transitions, along with uncertainties in the alignment
of polarization with respect to the magnetic field gradients, we
assign a 9 MHz uncertainty, along with a +4 MHz correction, to the
measured frequency.  Given this large uncertainty, we arrive at a
value of
\begin{align}
\nu(^3P_2 \rightarrow ^3S_1) = 486 \: 370 \: 098 (9) \: {\rm MHz}.
\end{align}
This value is an improvement by a factor of 50 over the current
value listed in Ref.\ \cite{sugar85} of $\nu(^3P_2 \rightarrow
^3S_1) = 486 \: 369 \: 853 (450)$ MHz.

We have performed a measurement of the ${\rm 4s4p}\: ^3P_1
\rightarrow {\rm 4s5s} \: ^3S_1$ transition in calcium using the
direct output of a femtosecond frequency comb.  This experiment
demonstrates the versatility of direct frequency-comb spectroscopy
by combining it with precision cw laser spectroscopy to achieve
high-precision absolute frequency measurements of transitions
between excited states at wavelengths that are otherwise difficult
to produce.

We thank J. Torgerson and Los Alamos National Laboratory (LANL) for
supporting the development of the 1 GHz Ti:Sapphire laser.
Additional funding came from the National Institute of Standards and
Technology. T.M. Fortier and J.E. Stalnaker acknowledge the support
of Director's funding from LANL and the National Research Council,
respectively.  This work was performed by NIST and is not subject to
U.S. copyright

\end{document}